\documentclass[amssymb,aps,twocolumn,showpacs, superscriptaddress,nofootinbib]{revtex4-1}
\usepackage[]{graphicx}
\usepackage[]{amsmath}
\usepackage{hyperref,color}

\def\ket#1{| #1 \rangle}

\def\kb#1#2{| #1 \rangle\!\langle #2 |}

\def\cC{\mathcal{C}}

\def\cE{\mathcal{E}}
\def\cF{\mathcal{F}}

\def\cU{\mathcal{U}}

\def\Tr{\mathrm{Tr}}

\def\eq#1{Eq.~\eqref{eq:#1}}
\def\fig#1{Fig.~\ref{fig:#1}}

\begin{document}

\title{Tensor Networks and Quantum Error Correction}
\author{Andrew J. Ferris}
\affiliation{ICFO---Institut de Ciencies Fotoniques, Parc Mediterrani de la Tecnologia, 08860 Barcelona, Spain}
\affiliation{Max-Planck-Institut f\"ur Quantenoptik, Hans-Kopfermann-Str. 1, 85748 Garching, Germany}
\affiliation{D\'epartement de Physique, Universit\'e de Sherbrooke, Qu\'ebec, Canada}
\author{David Poulin}
\affiliation{D\'epartement de Physique, Universit\'e de Sherbrooke, Qu\'ebec, Canada}

\date{\today}

\begin{abstract}
We establish several relations between quantum error correction (QEC) and tensor network (TN) methods of quantum many-body physics. We exhibit correspondences between well-known families of QEC codes and TNs, and demonstrate a formal equivalence between decoding a QEC code and contracting a TN. We build on this equivalence to propose a new family of quantum codes and decoding algorithms that generalize and improve upon quantum polar codes and successive cancellation decoding in a natural way. 
\end{abstract}


\maketitle

The basic principle of quantum error correction (QEC) is to encode information into the long-range correlations of entangled quantum many-body states in such a way that it cannot be accessed locally. When a local error affects the system, it leaves a detectable imprint---called the error syndrome. The decoding problem consists in inferring the recovery with greatest probability of success given the error syndrome. In general, this is a hard problem \cite{HL11a,IP13a}, but for well-chosen codes, it can be solved efficiently either exactly (e.g. \cite{Pou06b,PP12a}) or heuristically (e.g. \cite{PTO09a, DP10a}).  

Entangled many-body states generally have a number of parameters that increase exponentially with the number of particles, and so are not amenable to direct numerical calculations. Tensor network (TN) states were introduced  \cite{V03a,VC04a,Vid05a} as families of many-body states that are specified with only polynomially many parameters.
 In this setting, the evaluation of physical quantities of interest such as correlation functions and local expectation values reduces to contracting the indices of a TN. In general, the contraction of a TN is a difficult problem \cite{MS08a}, but some TNs can be efficiently contracted \cite{V03a,SDV06a,Vid05a,EV12b}, sometimes using heuristic approximations \cite{VC04a,GLW08a}. 

In this letter, we explore and deepen the relation between QEC and TNs and leverage this relationship to propose new encoding and decoding techniques. First, we establish a formal connection between the decoding problem for QEC  and TN contraction. We then describe the correspondence between a number of well known TN families and QEC codes.  Finally, we exploit this equivalence to propose a new family of efficiently decodable QEC codes that naturally generalize quantum polar codes \cite{WG13a,RDR11a,WG13b,RW12a,WLH13a,RSDR13a,DGW12a} based on a recently introduced family of TNs called branching MERA \cite{EV13a}. We study their performance numerically and find that they outperform polar codes. Further, our numerics clearly demonstrate good code performance is possible without using entanglement assistance. In a companion paper \cite{FP13}, we present a detailed study of the classical analogue of these new codes.  

\noindent{\em Quantum error correction---} In general, a $[[n,k,d]]$ quantum code $\cC$ is a $2^k$-dimensional subspace of an $n$-qubit Hilbert space. The integer $d$ is the minimum distance of the code, and indicates the number of simultaneous errors that it can correct. The code can be defined as the image of a unitary encoding circuit $U$ acting on an $n$-qubit state, where $k$ ``data qubits" can be in an arbitrary state and the other $n-k$ ``syndrome qubits" are restricted to the state $\ket 0$, 
\begin{equation}
\cC = \{ \ket\psi = U \ket\phi_k\otimes \ket 0^{\otimes n-k} : \ket\phi_k \in (\mathbb{C}^2)^{\otimes k}\}.
\label{eq:code}
\end{equation}
Subjected to an error $E$, the encoded state $\ket\psi$ is transformed to $\ket{\psi'} = E\ket\psi$. Measuring the syndrome qubits on the state $U^\dagger \ket{\psi'} $ yields the error syndrome, and the decoding problem consists in identifying the optimal recovery given the syndrome. In this letter, we define quantum codes in terms of their encoding circuit $U$.  

\noindent{\em Decoding as TN contraction---}
We now explain how the decoding problem can be expressed as a TN contraction. We focus on Clifford encoding circuits  and Pauli channels; the general case is treated in the Appendices. Recall that the $n$-qubit Pauli group consists of tensor products of the four Pauli matrices $I$, $\sigma_x$, $\sigma_y$, and $\sigma_z$ and that Clifford circuits map that group to itself.

Pauli channel noise models assign probabilities $P_n(E=E_1\otimes E_2\otimes\ldots\otimes E_n) = P(E_1)P(E_2)\ldots P(E_n)$ to each element $E$ of the Pauli group, where $P(E_i)$ is a probability distribution. Hence, $P$ is represented by a rank-one tensor of dimension 4, that is $P = (p_I,p_x,p_y,p_z)$.  Consider the distribution $Q_n(E) = P_n(U^{-1}EU)$ corresponding to the distribution of errors after the de-encoding circuit $U^{-1}$. This distribution is obtained by contracting the encoding circuit (viewed as a rank $2n$ tensor) with the $P$'s, as in \fig{decoding} (a). To decode, we condition this probability distribution on the observed error syndrome, and typically decode one qubit at a time for efficiency reasons.

Prior to encoding, the syndrome qubits were in state $\ket 0$, as in \eq{code}. After de-encoding, these qubits are measured in the basis $\sigma_z$ to reveal the error syndrome (the $\pm 1$ outcomes are usually denoted $\{0,1\}$).  A syndrome $0$ doesn't strictly imply an error-free qubit. Instead, it indicates that it either had no error or a $\sigma_z$ error---providing no information about the $z$-quadrature. Thus, the probability tensor representing such a measurement outcome is a bimodal indicator function on $I$ and $\sigma_z$, i.e. $b_z = (1,0,0,1)$. Similarly, a syndrome $1$  is consistent with either a $\sigma_x$ or a $\sigma_y$ error, corresponding to the bimodal indicator function $\bar b_z = (0,1,1,0)$. Note that $\overline b_z = \sigma_x b_z$. Thus, conditioning the distribution $Q_n(E)$ on the syndrome is achieved by contracting it with $b_z$ or $\overline b_z$.  To trace out a qubit from the distribution, we simply contract $Q_n$ with the uniform distribution ${\bf e} = (1,1,1,1)$ on that qubit. \fig{decoding} (a) illustrates the resulting TN. Later, to efficiently contract the TNs corresponding to the polar and branching-MERA codes, we will make use of circuit identities shown at \fig{decoding} (b-e). There, we also must prepare and measure qubits in the other quadrature. We use the bimodal indicator function $b_x = (1,1,0,0)$ 
in the case where a syndrome qubit prepared in the $\ket +$ state and later measured along $\sigma_x$.

We exhibit relations between well-studied QEC codes and TNs below:

\begin{figure}[t]
\includegraphics[width=8.5cm]{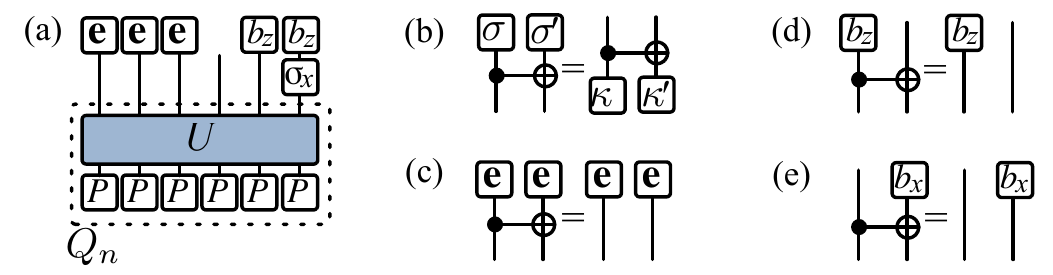}
\caption{(a) General decoding problem. This TN encodes the probability distribution $Q_4(E|s)$ of qubit 4 given that the 5th and 6th qubits revealed the syndrome $s=(0,1)$. (b-e) Action of the CNOT gate on the bimodal indicator functions. The truth table for (b) $(\sigma,\sigma') \rightarrow (\kappa,\kappa')$ is given by $(I,\sigma_x)\rightarrow (I,\sigma_x)$, $(\sigma_x,I)\rightarrow (\sigma_x,\sigma_x)$, $(I,\sigma_z)\rightarrow (\sigma_z,\sigma_z)$, and $(\sigma_z,I)\rightarrow (\sigma_z,I)$. Identities (c-e) follow from the application of (b) to bimodal indicator functions.}
\label{fig:decoding}
\end{figure}

\begin{figure*}[t!]
\includegraphics[width=17.5cm]{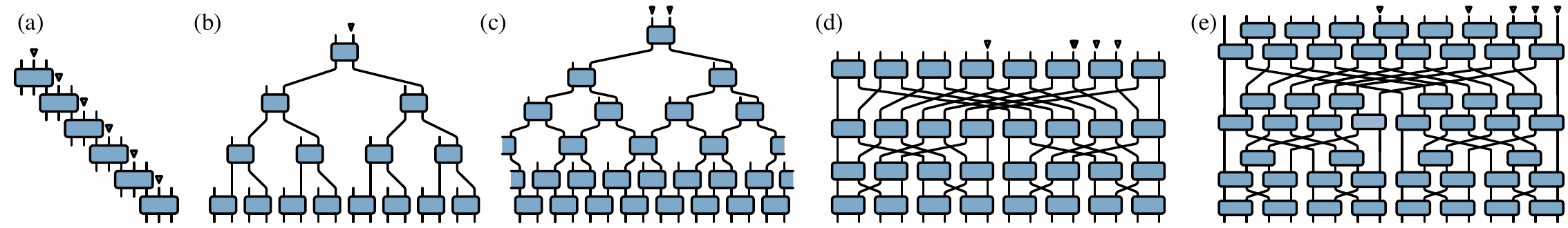}
\caption{Graphical definitions of various unitary TNs and encoding circuits of QEC codes: wires represent one or a few qubits, rectangles represent unitary transformation with time running from top to bottom. The size of each circuit can be varied in an obvious way. For coding applications, location of the data qubits are indicated by triangles, other qubits are initialized to $\ket 0$ (or $\ket +$).  (a) MPS TN and convolutional codes.   (b) Tree TN and concatenated quantum block codes. (c) MERA TN and topological code. (d) Branching tree TN and polar codes.  (e) Branching MERA TN and codes introduced in this letter.}
\label{fig:TN}
\end{figure*}

\noindent{\em Convolutional codes---} These codes are defined by the ``staircase" quantum circuit represented at \fig{TN} (a)  \cite{Cha98b,OT03a}. 
Given unentangled input data, the resulting encoded state will have rather limited entanglement, as entanglement between each step is mediated by only a constant number of ``memory" qubits. As a consequence, these codes have a constant minimal distance. The corresponding states define matrix product states (MPS) \cite{Orus2013}, which accurately describe the ground state of gapped one-dimensional quantum systems, and are the variational class of states behind density matrix renormalization group~\cite{Whi92a}.

It is possible to exactly evaluate the error probability of each data qubit using a message passing algorithm \cite{PTO09a}. This problem and corresponding method are formally equivalent to the transfer-operator method used to compute local expectation values from a MPS \cite{Orus2013}. In QEC, it is also possible to determine the globally optimal recovery 
 using Viterbi's algorithm \cite{PP12a}. Similar techniques have been proposed to determine the optimal MPS approximation to a ground state \cite{SC10a}.

\noindent{\em Turbo codes---} Turbo codes are constructed from the interleaved concatenation of convolutional codes, where the interleaver consists of a random permutation of the qubits between the two encoding circuits \cite{PTO09a}. This leads to long-range entanglement, and therefore larger minimum distances. The decoding procedure is analogous to an approximation used in many-body physics to solve two spin chains that are coupled by random, non-local interactions. In those terms, the transfer operator is used to evaluate local expectation values in each chain and the non-local inter-chain interactions are treated by mean field.   

\noindent{\em Concatenated block codes---} These are defined by the tree-shaped quantum circuit of \fig{TN} (b). Their importance stems from their role in fault-tolerance \cite{Sho96a}. Tree TNs underlie some early real-space renormalization methods for quantum systems \cite{K93a}, and continue to be used more recently \cite{SDV06a,Tagliacozzo2009,F13a, PVK13a}. The contractibility of this TN has led to an exact (maximum likelihood) decoding algorithm for concatenated codes \cite{Pou06b} offering significant improvements over conventional (minimum distance) decoders.

\noindent{\em Topological codes---} Such codes correspond to the degenerate ground space of a local, gapped two-dimensional Hamiltonian, which include model systems for topological order such as Kitaev's quantum double \cite{Kit03a} and Levin-Wen string-nets \cite{LW05b}. These systems can be described by projected entangled pair states (PEPS) \cite{SCP10a}, a family of TN that are a natural generalization of MPS to higher dimensions \cite{VC04a}. PEPS are generally not efficiently contractable, and consequently there exists no efficient, exact decoding algorithm for topological codes. Heuristic renormalization group methods have been devised both in the context of PEPS contraction \cite{GLW08a} and topological code decoding \cite{DP10a}, and in both case can offer accurate estimates.  

Topological codes can also be represented by a different family of TN --- the multi-scale entanglement renormalization ansatz (MERA) \cite{KRV09a}, defined at \fig{TN} (c).  These TNs accurately describe the ground states of critical one dimensional systems \cite{Vid05a}, as well as some two dimensional systems, including exact descriptions of some with topological order \cite{KRV09a}. MERA can be accurately contracted for the evaluation of $k$-point correlation function, but the complexity scales exponentially with $k$. Unfortunately, in this case the decoding problem is formally equivalent to the evaluation of an $n$-point correlation function, so the correspondence with MERA does not yield an efficient decoding method for topological codes.

\noindent{\em Polar codes---} The spectral (or branching tree) tensor network of \fig{TN} (d) has a structure identical to the fast Fourier transform and thus is useful for representing highly-entangled states of (non-interacting) fermions~\cite{F13b}. Arikan's polar code and its generalizations can be defined by that circuit, where every gate is a CNOT.  The exact location of the data qubits depend on the noise model and the rate of the code. Determining the optimal location of the data qubits (ignoring any correlations between decoding errors) is realized by Arikan's so-called `genie decoder' \cite{A09a}, and can be formulated as a tensor contraction problem. Decoding is realized using Arikan's successive cancellation decoder which similarly can be recast as a TN contraction as we will explain below. Because they obey a special self-duality condition, polar codes can be used to encode quantum information \cite{RDR11a,WG13b}, where in general they can attain the coherent information rate \cite{RW12a,RSDR13a}.

\noindent{\em Branching MERA codes---} The spectral TN can be seen as a simplification of branching MERA shown on \fig{TN}~(e), where half of the gates (sometimes called the disentanglers) have been removed. Branching MERA states were introduced in physics as a variational class of efficiently-contractible TN states with large amounts of entanglement \cite{EV12b}. While the amount of entanglement between a block of length $L$ and the rest of the chain is bounded by a constant in MPS and by $\log L$ in MERA and tree TN, it can be as large as $L$ in a branching MERA, the maximum allowed by quantum mechanics. 

We define {\em branching MERA codes} to be the (classical or quantum) code resulting from the family of encoding circuit illustrated \fig{TN} (e), with all gates CNOTs, and with the location of data qubits depending on the channel and encoding rate. (Note that, unlike the tree and MERA codes mentioned earlier, the orientation of this unitary circuit is reversed compared to their usual orientation in many-body physics, making the contractions efficient). One small difference in the quantum setting is that not all input stabilizer qubits are set to the state $\ket 0$; some are set to $\ket +$. The choice of which input qubit carries data, which are set to $\ket 0$ and which are set to $\ket +$ is a problem called channel selection which will be addressed below. Before, we explain how decoding is done.

\begin{figure*}[]
\includegraphics[width=18cm]{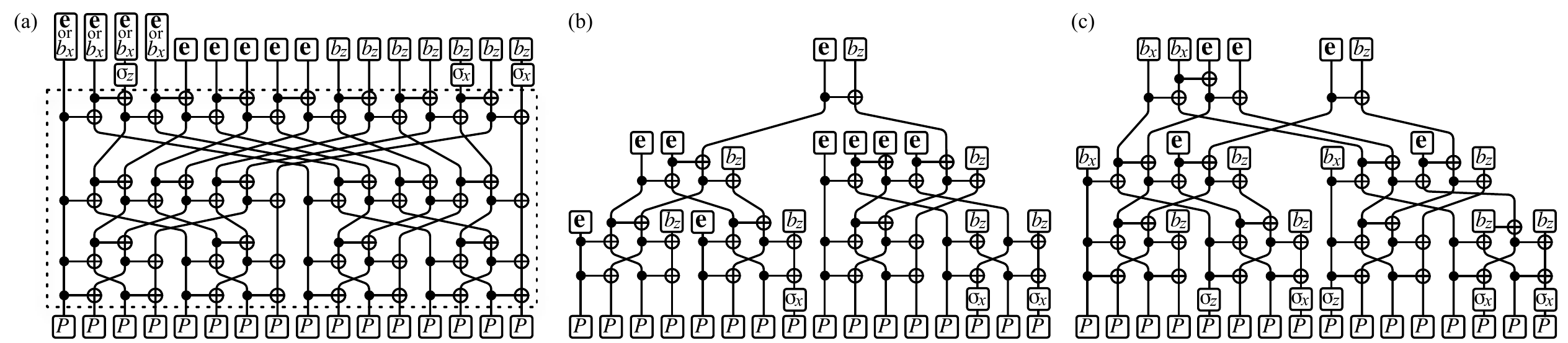}
\caption{Tensor network representation of the branching MERA code. (a) This TN encodes the $x$ error probability of the 7th qubit (from right) knowing that among the first 6 qubits, only the 1st and 3rd had $x$ errors, among the last 4 qubits, only qubit 14 had a $z$ error, and ignoring everything about qubits 8 to 12. The decoder can choose to ignore the information about $x$ errors, in which case the 4 tensors $b_x$ on the left are replaced by $\bf e$. Using the circuit identities of \fig{decoding}, these two TNs for $z$ decoder and the symmetric $x$-$z$ decoder become equivalent to (b) and (c).}
\label{fig:Decoder_bMERA}
\end{figure*} 

\noindent{\em Decoding polar and branching MERA codes---}
We first adapt Arikan's successive cancellation decoder to the quantum setting by decoding the two quadratures ($x$ and $z$ errors) in succession. First, $x$ errors are decoded sweeping from right to left. Assume that all qubits to the right of position $i$ have a fixed bimodal distribution $b_z$ or $\overline b_z$, either because they were $z$-type syndrome qubits or because we have already determined if they have undergone $x$-type error. The idea of successive cancellation is to determine the value of $x$-type errors on qubit $i$ conditioned on this information only, ignoring any information about qubits to the left of position $i$, which are thus contracted with $\bf e$. The resulting TN for branching-MERA is therefore represented by \fig{Decoder_bMERA} (a). Using the circuit identities of \fig{decoding}, this TN is equivalent to \fig{Decoder_bMERA} (b), which has a tree-width 3 (meaning that the intermediate steps of the decoder need only deal with distributions over at most 3 qubits), and can therefore be efficiently contracted (with cost $\mathcal{O}(n)$). Afterwards, $z$-type errors are decoded in a similar fashion, sweeping from left to right reducing the bimodal distributions $b_z$ and $\overline b_z$ to a single Pauli channel ($I$, $\sigma_x$, $\sigma_y$, $\sigma_z$). When taking care to recycle previous calculations, the total decoding procedure has numerical cost $\mathcal{O}(n\log n)$. Decoding of polar codes follows straightforwardly since they are obtained by removing gates from branching MERA circuits.

We have also implemented a symmetric decoder where $x$- and $z$-type errors are decoded simultaneously, sweeping from the right for  $x$-errors and from the left for $z$-errors. This can provide an advantage because $x$ and $z$ errors are typically correlated. On a depolarization channel for instance, where $p_x=p_y=p_z$, this correlation is mediated by $y$ errors which are seen as a combination of $x$ and $z$ errors (remember $\sigma_z\sigma_x = i\sigma_y$). Thus, knowing about $z$ errors at the time of decoding $x$ errors provides additional information which enhances the decoding performance. As shown at \fig{Decoder_bMERA}~(c), the resulting TN has tree-width equal to 6 for branching-MERA (or 2 for polar codes), so this decoding scheme is also efficient. 

To understand the origin of this contractibility, and hence the efficiency of the two decoders described above, note that the circuit identities can be used to remove all CNOTs in the encoding circuit \fig{Decoder_bMERA} (a) when all top tensors are identical (either all $\bf e$, $b_x$, $b_z$ or $I$). Thus, the complexity comes from ``domain walls" between different kinds of distributions. Contracting a TN with 1 or 2 domain walls is equivalent to computing a 1 or 2-point correlation of a branching MERA, which can be done efficiently \cite{EV13a}.

\noindent{\em Numerical results---}
We now numerically investigate the performance of these codes and decoding techniques for the depolarizing channel (further results, including for the erasure channel, are in the Appendices). 

\begin{figure}[t]
\includegraphics[width=0.5\columnwidth]{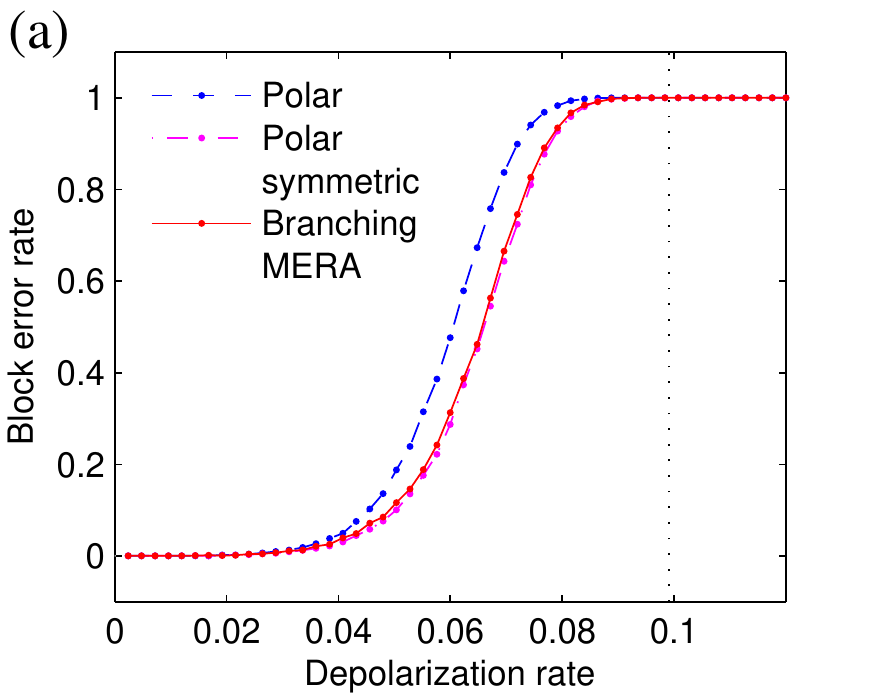}\includegraphics[width=0.5\columnwidth]{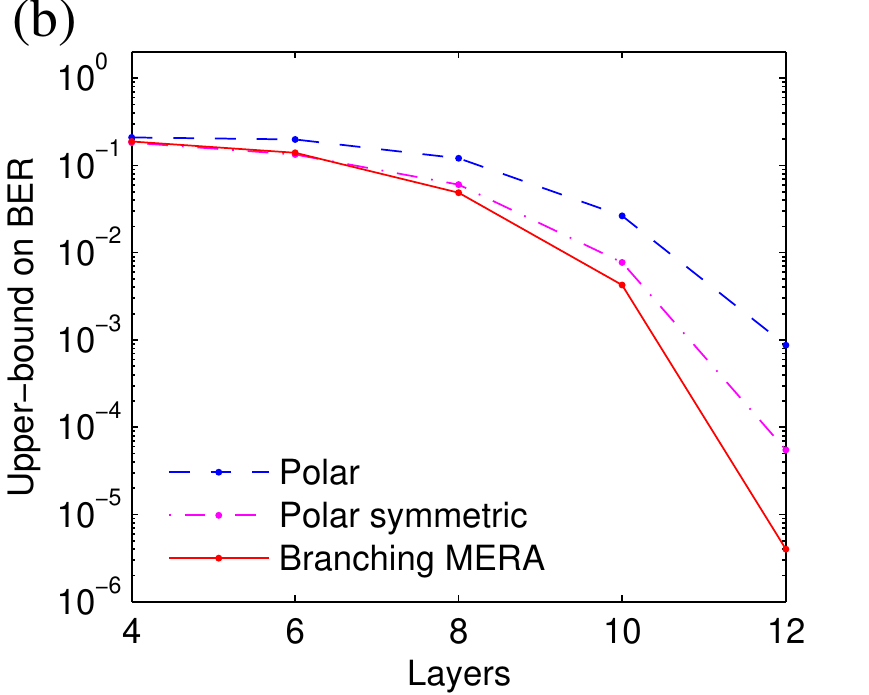}
\caption{Performance of the polar and branching-MERA codes under the depolarizing channel. The block-error rate (BER) is the probability of incorrectly decoding a block on $n$ qubits. (a) Comparison of the various codes and decoding algorithms for codes of size $2^{12}$ qubits with encoding rate 1/2. 
(b) Upper bound on the BER for codes with depolarizing rate 9.92\% and encoding-rate 1/8 as a function of system size. The probability of error decreases strongly with code size. Both the symmetric decoder and the branching-MERA improve beyond the standard quantum polar code.}
\label{fig:qdresults}
\end{figure} 

The first stage of the procedure is to select which logical channels will cary quantum information and which will be fixed in either the $x$ or $z$ basis. The performance of a single quadrature of a given logical channel can be characterized independently from the remainder with Monte Carlo sampling techniques. In this procedure, we determine the probability that a given decision (i.e. determining an error in either the $x$ or $z$ basis) is incorrect \emph{assuming all the prior decisions were correct}. As in \cite{DGW12a}, we then select the data channels to be those with the lowest error rate on the worst quadrature (and fixing the remaining qubits in the $x$ and $z$ basis depending on which quadrature performs worse). See the Appendices for a detailed study of channel polarization.

In \fig{qdresults}~(a), we plot the performance of the polar code (with standard decoder), the polar code with symmetric decoder and the branching-MERA code (with standard decoder) as a function of depolarization probability. In all cases we see a relatively sharp crossover between a low-error rate regime and a high-error rate regime, occurring somewhat below the depolarizing rate 9.92\% where the coherent information and code rates coincide.  We observe that the threshold approaches the capacity with increasing code size (although, much like is observed for the classical polar code, this approach is relatively slow), and that both the improved symmetric decoder and branching-MERA code have a sharper transition at equal block size. Thus, both of these improve what is the main drawback of polar codes, which is their important finite-size effects.

In \fig{qdresults}~(b) we study the performance as a function of code-size. These results are a simple upper-bound on the error rate achieved by summing the individual error rates of the data channels (both quadratures) and the frozen channels (the non-frozen quadrature) that were generated in the channel-selection phase. We observe in all cases that the block-error rate decreases rapidly with the number of layers used in the codes. For classical polar codes it is known that the error rate scales as $\exp(-\sqrt{n})$, however we would require more computational resources to study this scaling here. 

Finally, our numerical results provide additional evidence that entanglement assistance is unnecessary for achieving good performance with quantum polar codes (or the branching-MERA code). Entanglement assistance was proposed~\cite{RDR11a,WG13b} to address possible channels that exhibit poor performance in both quadratures. Fortunately, channel polarization ensures that performance is good in either or both quadratures, and further we observe that polarization improves rapidly with code-size. We make this more precise in the Appendices.

\noindent{\em Conclusion---} 
In summary, we have found that TNs provide a powerful description of QEC codes and decoding algorithms, leading us to suggest new encoding and decoding techniques beyond the standard polar code. We observe that these achieve very-low error rates at high data rates (approaching the coherent information rate), especially for large code sizes. The encoding and decoding procedures follow closely their classical counterparts (by treating each quadrature in succession) and the code performance provides additional evidence that polar codes require no pre-shared entanglement.

\noindent{\em Acknowledgements---}
This research is funded by Canada's NSERC and FQRNT through the network INTRIQ, as well as TOQATA (Spanish grant PHY008-00784), the EU IP SIQS, and the MPI-ICFO collaboration. Numerical resources were provided by Compute Canada and Calcul Qu\'ebec.

\bibliographystyle{apsrev}
\bibliography{../qubib}

\appendix

\section{General decoding problem}

In the main text, we have demonstrated that decoding a QEC code is formally equivalent to contracting a TN in the case where the encoding circuit is a Clifford transformation and the noise is modelled by a memoryless probability distribution over Pauli operators. Here, we extend this equivalence to arbitrary encoding circuits and memoryless noise models. 

There are many inequivalent formulations of decoding that have been implemented in different settings. For a code encoding $k$ data qubits into $n$ physical qubits, we may want to determine optimal way of recovering the $i$th data qubit, which we could name a marginal logical decoder. Alternatively, we may want to optimally recover the $i$th physical qubit, corresponding to a marginal physical decoder. For instance, both of these marginal decoders can be realized efficiently for convolutional codes \cite{PTO09a}, and there exists a heuristic method for the latter for LDPC codes \cite{PC08a}. Instead of marginal decoding, we may be interested in the global optimal recovery. We may ask for the globally most likely error affecting the physical qubits or the most likely error affecting the logical qubits. These two concepts differ in the quantum setting due to a phenomenon called error degeneracy  \cite{WHP03a,IP13a,PP12a}. 

Here, we will illustrate the principle for optimal marginal decoding of a data qubit, as we did at Fig. 1 (a) of the Main Text in the case of Clifford encoding with Pauli noise. We encode $k$ data qubits into $n$ qubits using an encoding circuit $U$. The encoded qubits are then subject to a memoryless channel $\cE = \cE_1\otimes \cE_2\otimes\ldots\otimes\cE_n$ where each $\cE_j$ is a completely-positive trace-preserving (CPTP) map. The encoding is undone using $U^\dagger$, and the syndrome qubits are measured to yield the error syndrome $s\in \{0,1\}^{n-k}$. Our goal is to determine the CPTP map to be applied to data qubit $j$ in order to maximize its average fidelity to its input value. Thus, it suffices to determine the effective channel on data qubit $j$ resulting from the above procedure.

\begin{figure}[t]
\includegraphics[width=4cm]{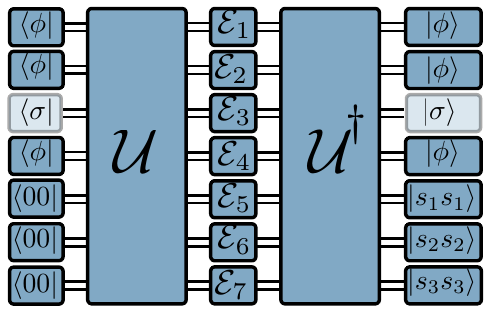}
\caption{Ignoring the shaded gates, TN representation of the  effective channel $\cF_{2|s}$ on second data qubit conditioned on the syndrome $s=s_1s_2s_3$. Gates in this circuits are CP maps, so each wire is doubled to carry a physical qubit and its time-inverted self. The encoding gate is $\cU = U\otimes U^*$, ket $\ket{s_is_i}$ represents the projector $\kb{s_i}{s_i}$, ket $\ket\phi = \ket{00}+\ket{11}$ represents a partial trace operation, and its dual bra represents the (unnormalized) maximally mixed state $I$. With the shaded gates added, circuit encodes $\Tr [\cF_{j|s}(\sigma) \sigma]$.}
\label{fig:FT}
\end{figure}

Denote this channel $\cF_{j|s}$. It is a four-legged tensor which is represented at \fig{FT}. Because we can determine $\cF_{j|s}$ given $\Tr [\cF_{j|s}(\sigma) \sigma]$ for a set of $\sigma$ for which $\sigma\otimes\sigma^*$ span a basis, we can cap the open wires by $\ket\sigma$ on both ends of the circuits. The resulting circuit corresponds to a TN for the expectation value $\langle \cE_1\cE_2\ldots\cE_n\rangle$, so we see that evaluating the effective conditional channel $\cF_{j|s}$---that consists the difficult step of decoding---is reduced to the evaluation of an $n$-point correlation function of a TN state. Depending on the nature of the code, this TN can be a MPS, MERA, tree, spectral tensor network (like the FFT or polar code), etc., and TN contractions schemes can be leveraged to decode the corresponding code.

\section{Channel polarization, the erasure channel and entanglement assistance}

Here we present additional numerical results on channel polarization, a study of the erasure channel similar to that of the depolarizing channel in the main text, and discuss entanglement assistance for the polar and branching-MERA codes.

\begin{figure*}[t!]
\begin{centering}
\includegraphics[width=0.5\columnwidth]{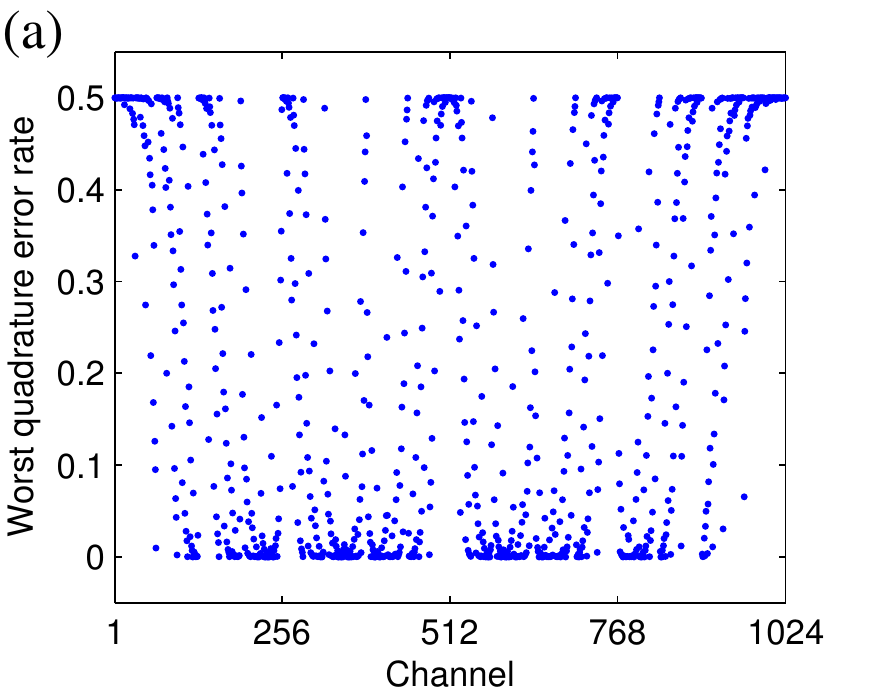} \includegraphics[width=0.5\columnwidth]{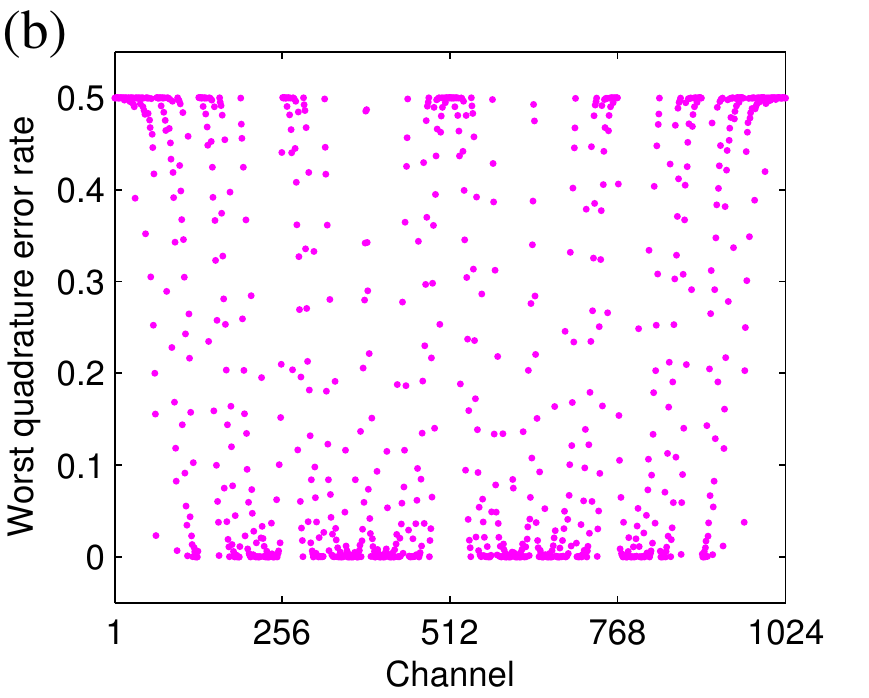} \includegraphics[width=0.5\columnwidth]{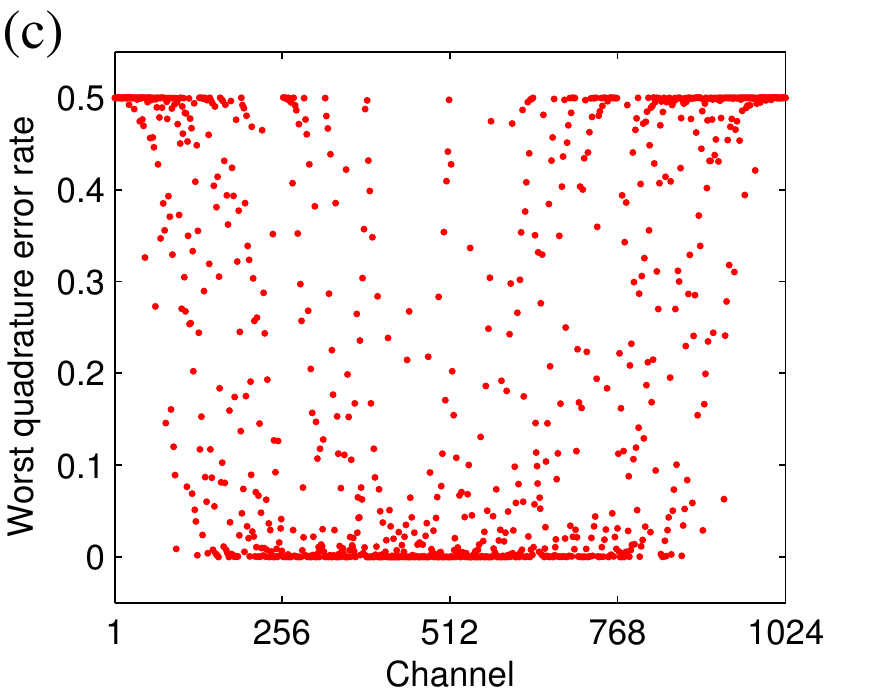} \\
\includegraphics[width=0.5\columnwidth]{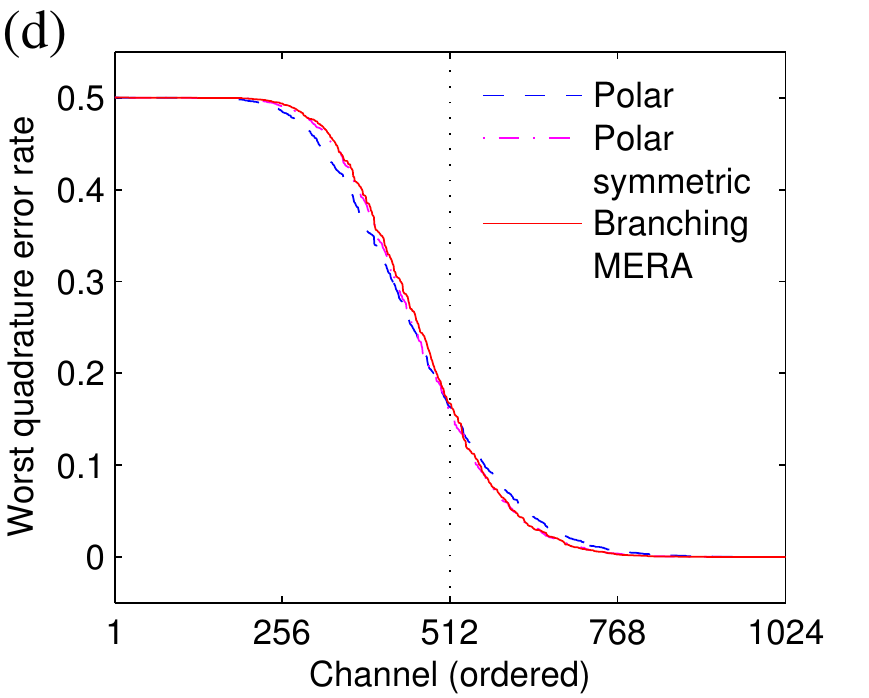} \includegraphics[width=0.5\columnwidth]{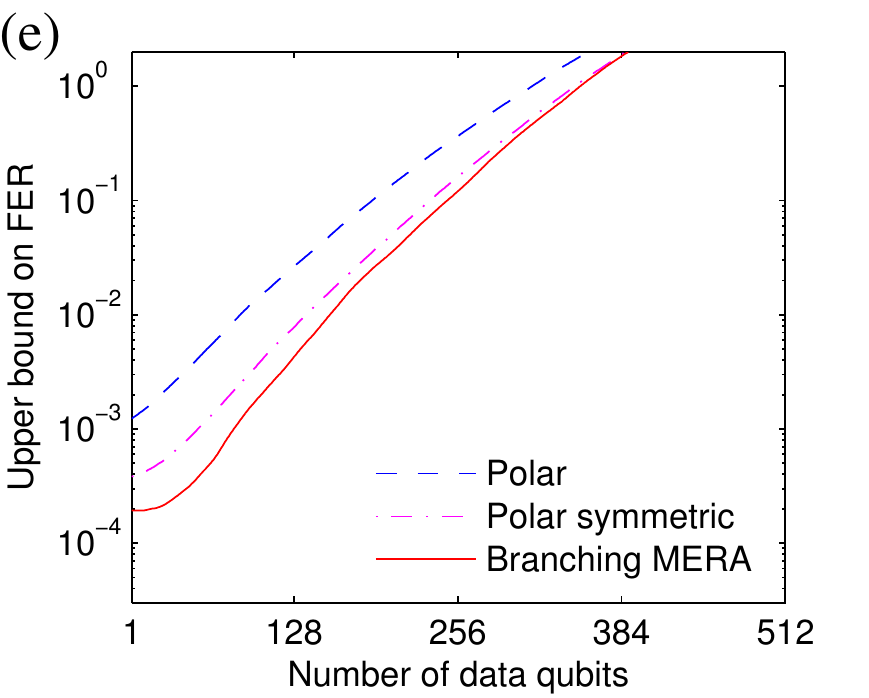}
\end{centering}
\caption{Channel polarization of the polar code and branching-MERA codes containing 1024 qubits under a 9.9\% depolarizing channel (capacity 50\%). In (a--c) we plot the value of the worst-quadrature error for each channel, using (a) polar code with standard decoder, (b) polar code with symmetric decoder, and (c) branching-MERA code with standard decoder. These results are plotted in descending order in (d). The upper-bound to the frame-error rate is given in (e) for the given number of data qubits. }
\label{fig:qdchannel}
\end{figure*} 

Channel polarization is an effect that is defined with respect to an (impractical) decoding algorithm using the so-called `genie decoder'~\cite{A09a}. In this scheme, each sub-channel (i.e. a single quadrature of a given qubit) is studied individually by making the rather strong assumption that all the previous decoding steps were successful (there were no prior errors). In the classical polar and branching-MERA codes, this is equivalent to determining the error rate of a particular logical channel given that every other channel is frozen. The situation is more subtle in the quantum case, because only a single quadrature can be frozen on each logical channel. Nonetheless, the genie decoder can be implemented efficiently in numerical Monte Carlo experiments (given the decoder has knowledge of the actual error). 

Much like in the classical case, these channel error rates can be summed to produce an overestimate of the frame-error rate (the probability of an error occurring is upper-bounded by the sum the of probabilities that a given decision is incorrect given the earlier decisions were correct). In this argument, we have assumed that degenerate errors may lead to a data error. Degenerate errors are defined as those that leave the encoded quantum state unchanged, such as applying the $\sigma_z$ operator to a channel frozen to state $|0\rangle$. Although such an error does not affect the transmitted quantum information (and thus one might assume does not lead to decoding errors), the successive-cancelation decoder will use this (incorrect) information when making future decisions. We have observed that any incorrect decision will act to scramble the remaining decoding steps --- that is, any error (even a degenerate one) is catastrophic. Events where the decoder correctly recovers the data qubits but incorrectly determines the degenerate errors are extremely rare in our simulations involving large data rates. Below we will observe that for very small data rates the degenerate errors become relatively more frequent, but for fixed error-rate with increasing code-size the probability of even a single degenerate error decreases very rapidly (because the stabilizers have a large weight by construction). This suggests to us that the quantum polar and branching-MERA codes presented here are non-degenerate codes and thus can not achieve good performance beyond the coherent information (Hashing bound), which is a different point of view than expressed in \cite{RSDR13a}, albeit for a slightly different polar coding scheme. 

In \fig{qdchannel} we present the channel polarization results for the depolarizing channel with code-size $2^{10}$ and depolarizing rate 9.92\% (corresponding to a symmetric capacity of 1/2). The data is produced with $10^7$ Monte Carlo samples using the genie decoder, and \fig{qdchannel} (a--c) display the performance of the worst quadrature of each qubit. In all cases we observe a strong channel polarization effect --- most of the channels have error rates close to 0 (perfect) or 0.5 (scrambled). In \fig{qdchannel}~(d) we observe that the polarization effect is slightly stronger for the improved symmetric decoder and for the branching-MERA code, compared to the standard quantum polar code. In all cases there is a tendency for to see large $z$-error rates on the left, and large $x$-error rates on the right, and this splitting effect is slightly stronger for the branching-MERA code. This separation has the beneficial effect of giving the successive-cancellation decoder access to more of the syndrome measurements when performing the early stages of decoding (and therefore allowing the decoder to perform closer to the optimal, maximum-likelihood decoder).

In \fig{qdchannel}~(e) we plot an upper-bound on the frame-error rate as a function of the number of data qubits. This is achieved by summing the error rates of both quadratures of the data qubits and the unprotected quadrature of the syndrome bits. We observe an error floor at very low encoding rates that can be explained by the fact that we are including all degenerate errors. Of course, these results are an upper bound only and actual performance may be somewhat better than this. We discuss degenerate errors and their relationship to entanglement assistance in more detail below.

\begin{figure}[t!]
\includegraphics[width=0.5\columnwidth]{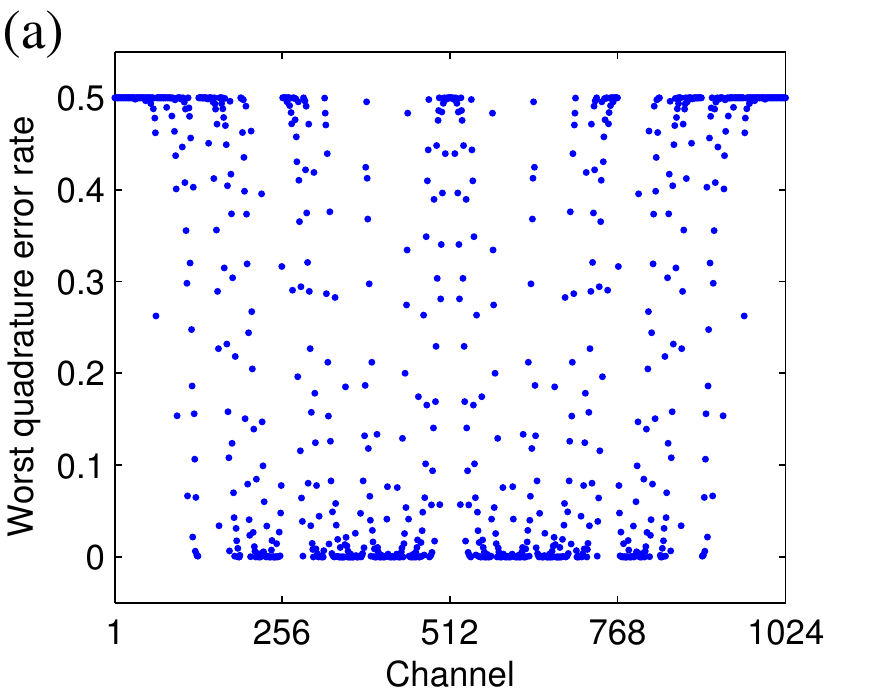}\includegraphics[width=0.5\columnwidth]{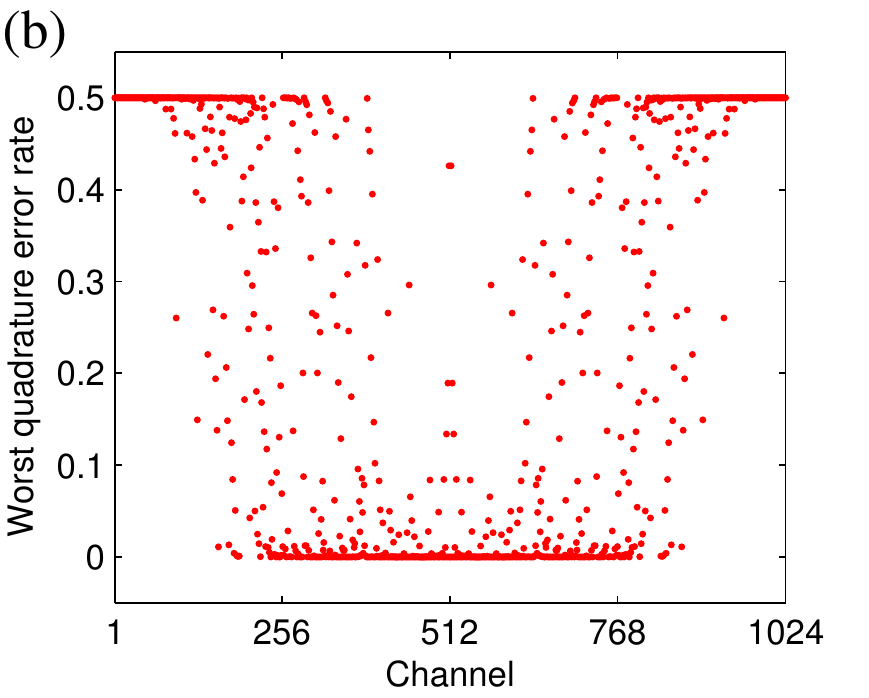}
\includegraphics[width=0.5\columnwidth]{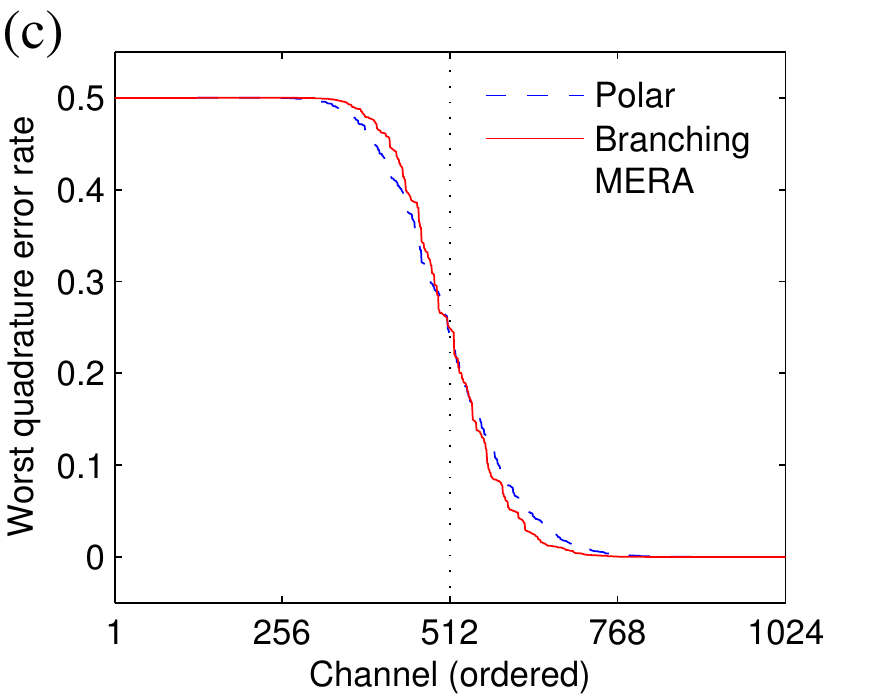}\includegraphics[width=0.5\columnwidth]{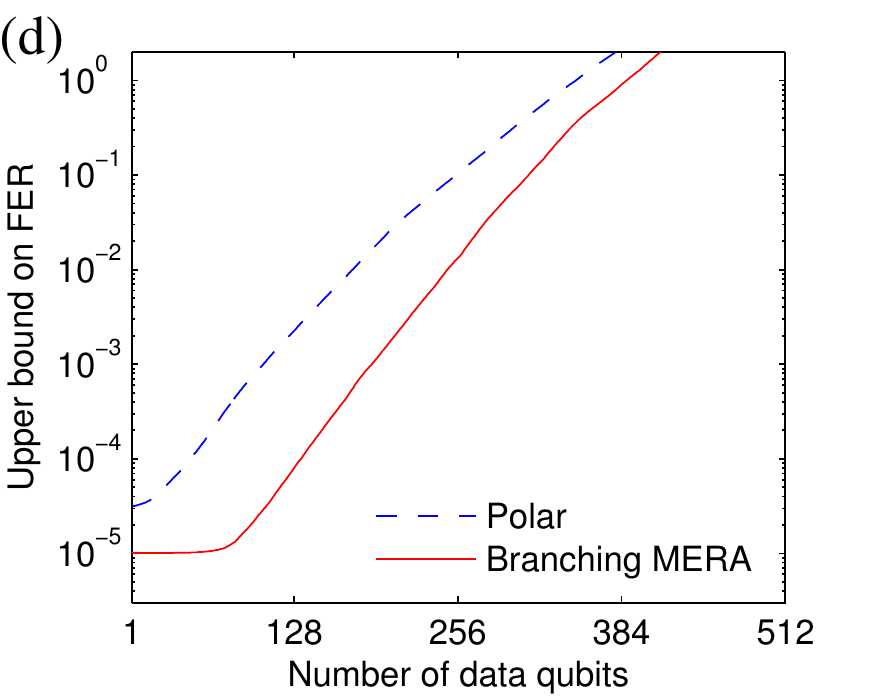}
\caption{Channel polarization of the polar code and branching-MERA codes containing 1024 qubits under a 25\% quantum erasure channel. In (a,b) we plot the value of the worst-quadrature error for each channel, for (a) the polar code and (b) branching-MERA code. These are plotted descending order in (c). The upper-bound to the frame-error rate is given in (d) for the given number of data qubits. }
\label{fig:qechannel}
\end{figure} 

We have performed a similar analysis for the quantum erasure channel in \fig{qechannel}. The quantum erasure channel has two qualitative differences to the depolarizing channel. Firstly, we can study the channel performance \emph{exactly} by keeping track of  a finite range of possible states of knowledge. For instance, the presence of an $x$-error on a single qubit may only be either fully determined or undetermined based on the information available. For two or more qubits, we have to include correlation effects such as knowing the sum of two values, but not the value of either. These correlations are essential for calculating the performance of the branching-MERA code (which deals with states of three qubit channels in the decoding process). The second difference is that for any channel with decoupled $x$ and $z$ errors, the performance of each quadrature is given independently (and in strong analogy with the classical case). Although the location of the errors in the erasure channel are correlated by their positions (a given qubit is erased or not), knowledge of the presence of an $x$-error does not help determine the pattern of $z$-errors because the erasure locations are known to the decoder at all times. This leads to the symmetry observed in \fig{qechannel}~(a,b), and further means that the symmetric decoder performs identically to the standard decoder. Otherwise, the results are qualitatively similar to that of the depolarizing channel, including the observation of an error-floor due to degenerate errors.

\begin{figure}[t!]
\vspace{4cm}
\includegraphics[width=0.5\columnwidth]{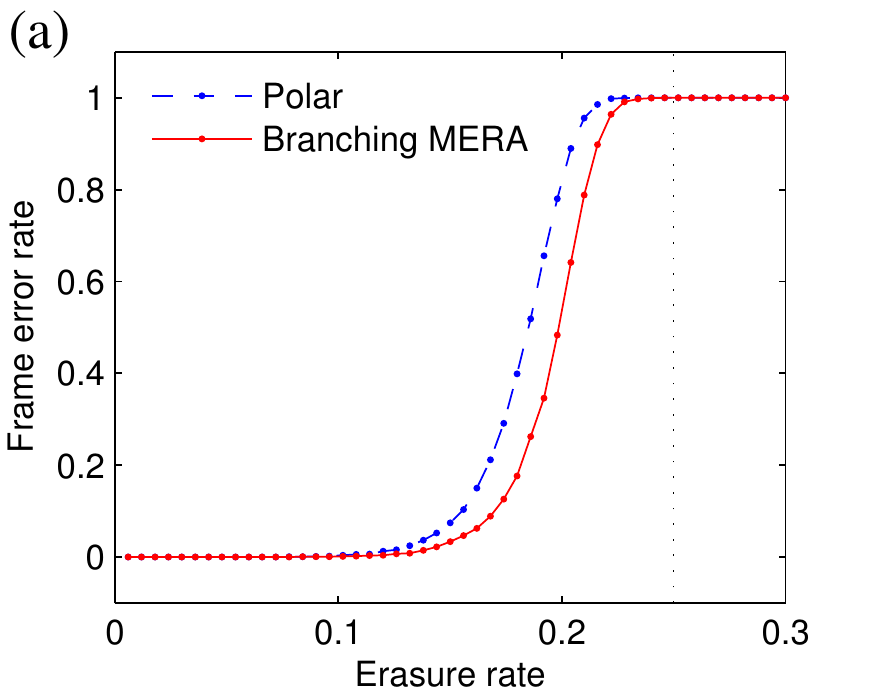}\includegraphics[width=0.5\columnwidth]{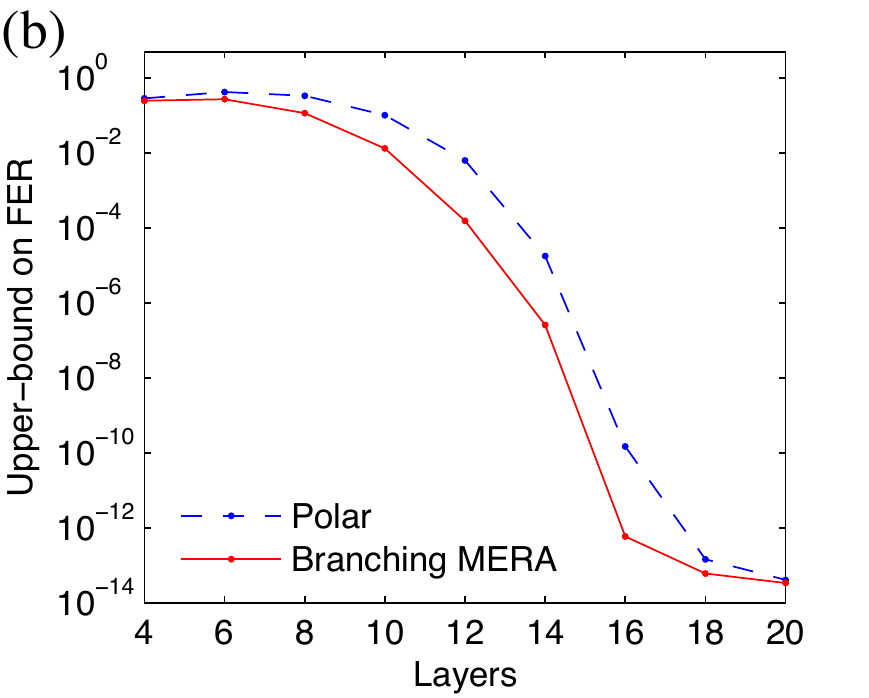}
\caption{Performance of the polar and branching-MERA codes under the quantum erasure channel. (a) Comparison of the two schemes for codes of size $2^{12}$ qubits with encoding rate 1/2. We observe that as the code-size increases, the `waterfall' region approaches the capacity corresponding to a erasure rate of 25\%. (b) Upper bound on the frame-error rate for codes with erasure rate 25\% and encoding-rate 1/4 as a function of system size. The probability of error decreases very strongly with code size.}
\label{fig:qeresults}
\end{figure} 

In \fig{qeresults}, we plot the performance of the codes under the quantum erasure channel in analogy to the results in Fig. 4 of the main text for the depolarizing channel. Generally, we observe that finite-size effects are reduced for the erasure channel (the waterfall is closer to capacity for a given code-size). In \fig{qeresults}~(b) the exact nature of the calculations allow us to reach much larger code-sizes and smaller error rates than for the depolarizing channel. For encoding rates of 1/4 and erasure rates of 25\% (the encoding rate is half of capacity), we see that the frame-error rate is reduced to a very low level of below $10^{-12}$ already for $2^{16}$ qubits --- which we believe is an impressive result for a quantum code. Further, we expect the performance under the depolarizing channel to improve similarly with code-size.

\begin{figure}[t!]
\vspace{4cm}
\includegraphics[width=0.5\columnwidth]{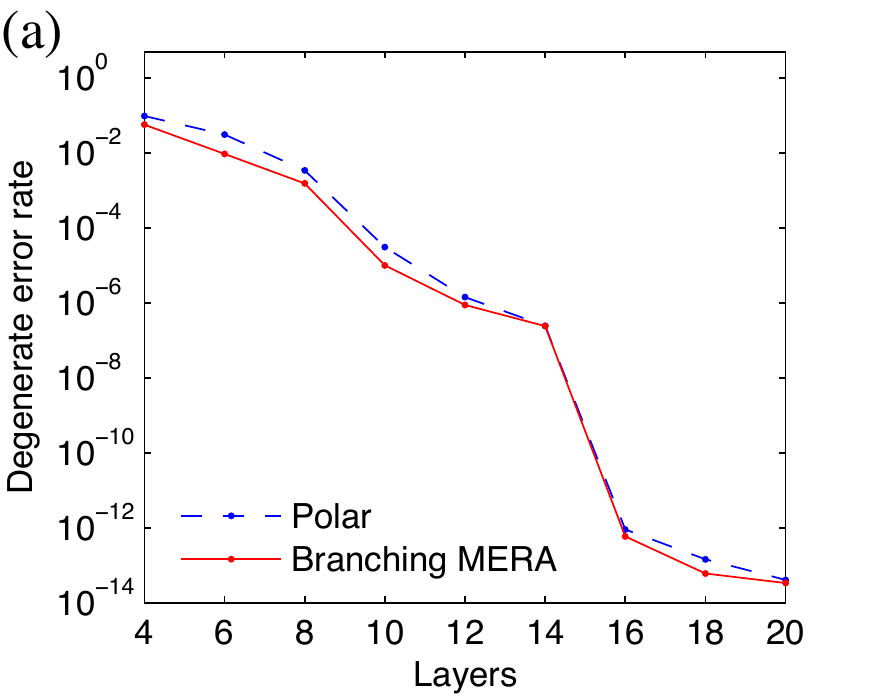}\includegraphics[width=0.5\columnwidth]{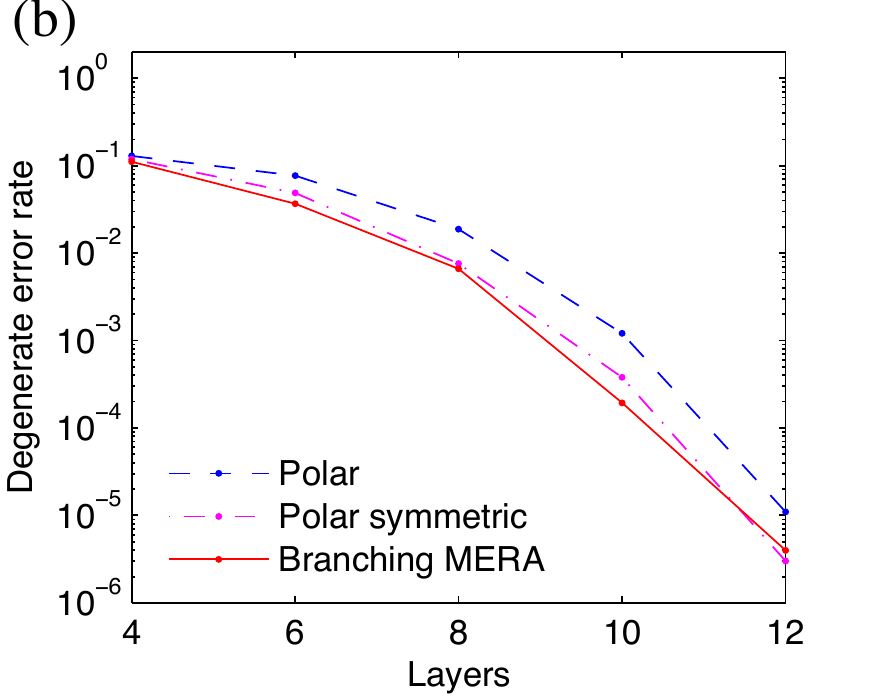}
\caption{Upper bounds for the probability of a degenerate decoding error in the limit of zero encoded qubits, for (a) the 25\% erasure channel and (b) the 9.92\% depolarizing channel. We observe that such errors decay extremely quickly with code size, indicating that channel polarization ensures that at least one quadrature becomes 'perfect' for large codes.. This obviates any for need for entanglement assistance for successfully decoding either the polar or branching-MERA code. }
\label{fig:qdegenerate}
\end{figure} 

Finally, we return to the issue of degenerate errors and entanglement assistance. Both Renes, Dupuis and Renner~\cite{RDR11a} and Wilde and Guha~\cite{WG13b} have proposed that entanglement assistance to be a possible way of dealing with qubits that have poor performance in both quadratures. This would work by combining such a faulty qubit with a well-behaved one so that errors in both quadratures can be measured. Further work~\cite{RW12a,RSDR13a} suggests entanglement assistance may not be necessary, possibly by using a more elaborate encoding and decoding scheme.

Our numerical study demonstrates that the likelihood of channel-errors in both quadratures of a given qubit diminishes very quickly with code-size, obviating the need for complex entanglement assistance schemes. For channels with independent $x$ and $z$ errors (including the erasure channel) this result is straightforward to demonstrate. For independent quadrature quantum noise models where the quadrature error rates coincide, we have $P_{z}(i) = P_{x}(n-i-1)$. Channel polarization of classical polar codes therefore demonstrates that (for sufficiently large code-sizes) the better-behaving quadrature of a given qubit polarizes to a perfect channel so long as the coherent quantum capacity is positive. For channels with correlations between $x$ and $z$ errors the argument is less forthcoming (see \cite{RW12a}), but numeric evidence displays similar behavior as for the independent noise models.  In \fig{qdegenerate}, we display the sum of the best-quadrature error rates (forming an upper bound to the rate of degenerate errors in the limit of zero encoded qubits). In both cases this rate decreases rapidly with the number of layers in the polar or branching-MERA code --- and the result is particularly clear for the erasure channel where exact calculations allow us to study larger code sizes. However, for the depolarizing channel we cannot rule out definitively the possibility of poor performance when the coherent information rate is very small, nor the possibility of a qualitative difference in the performance of the symmetric and standard decoding schemes in such a regime. We have also not tried to investigate performance beyond the coherent information limit, as discussed in \cite{RSDR13a}.

\end{document}